\documentclass[aps,prm,twocolumn,superscriptaddress,amsmath,amssymb,revsymb,showpacs]{revtex4-2}

\usepackage{color}
\usepackage{graphicx}
\usepackage{dcolumn}
\usepackage{bm}
\usepackage{url}
\usepackage{here}
\usepackage{color}
\usepackage{CJKutf8}

\newcommand{\be}{\begin{eqnarray}}
\newcommand{\ee}{\end{eqnarray}}

\newcommand{\nn}{\nonumber}

\newcommand{\bo}{\boldsymbol}
\newcommand{\lb}{\label}
\begin{document}


\title{Specific Heat Anomalies and Local Symmetry Breaking in (Anti-)Fluorite Materials: A Machine Learning Molecular Dynamics Study}

\author{Keita Kobayashi}%
\email{kobayashi.keita@jaea.go.jp}
\affiliation{%
CCSE, Japan Atomic Energy Agency, 178-4-4, Wakashiba , Kashiwa, Chiba
 277-0871, Japan}%

\author{Hiroki Nakamura}%
\affiliation{%
CCSE, Japan Atomic Energy Agency, 178-4-4, Wakashiba , Kashiwa, Chiba
 277-0871, Japan}%

\author{Masahiko Okumura}%
\affiliation{%
CCSE, Japan Atomic Energy Agency, 178-4-4, Wakashiba , Kashiwa, Chiba
 277-0871, Japan}%

\author{Mitsuhiro Itakura}%
\affiliation{%
CCSE, Japan Atomic Energy Agency, 178-4-4, Wakashiba , Kashiwa, Chiba
 277-0871, Japan}%

\author{Masahiko Machida}%
\affiliation{%
CCSE, Japan Atomic Energy Agency, 178-4-4, Wakashiba , Kashiwa, Chiba
 277-0871, Japan}%

\date{\today}

\begin{abstract}
Understanding the high-temperature properties of materials with (anti-)fluorite structures is crucial for their application in nuclear reactors. In this study, we employ machine learning molecular dynamics (MLMD) simulations to investigate the high-temperature thermal properties of thorium dioxide, which has a fluorite structure, and lithium oxide, which has an anti-fluorite structure. Our results show that MLMD simulations effectively reproduce the reported thermal properties of these materials. A central focus of this work is the analysis of specific heat anomalies in these materials at high temperatures, commonly referred to as Bredig, pre-melting, or $\lambda$-transitions. We demonstrate that a local order parameter, analogous to those used to describe liquid-liquid transitions in supercooled water and liquid silica, can effectively characterize these specific heat anomalies. The local order parameter identifies two distinct types of defective structures: lattice defect-like and liquid-like local structures. Above the transition temperature, liquid-like local structures predominate, and the sub-lattice character of mobile atoms disappears. 
\end{abstract}

\maketitle
\section{Introduction}

Uranium, plutonium, and thorium dioxides (UO$_2$, PuO$_2$, and ThO$_2$), which exhibit fluorite structures, are representative nuclear fuel materials. 
Lithium oxide (Li$_2$O), with anti-fluorite structure, is a candidate material for tritium breeding for nuclear fusion reactor \cite{van2000ceramic}. 
In both cases, understanding the material properties over a wide range of temperatures is essential for reactor design and ensuring nuclear safety.

Atomic simulations, such as molecular dynamics (MD) simulations \cite{potashnikov2011high, CRG, cooper2014thermophysical, galvin2016thermophysical, hayoun2005complex, oda2009modeling, Asahi_2014, lavrentiev2019lithium} and density functional theory (DFT) calculations \cite{f-electon, szpunar2014theoretical, szpunar2016theoretical, nakamura2016high, gupta2019lithium, jaberi2024study}, are effective tools for complementing experimental data and unraveling phenomena that remain unresolved at the atomic level.
Among these methods, machine learning molecular dynamics (MLMD) simulations \cite{BPNN1, GAP1, BPNN2, GAP2} are particularly well-suited for investigating the high-temperature properties of materials.
In MLMD, flexible functions with multiple adjustable parameters, such as artificial neural networks \cite{BPNN1,BPNN2} and Gaussian processes \cite{GAP1,GAP2}, are employed as the interatomic potentials for MD simulations.
These interatomic potentials, trained to imitate the DFT potential energy surface accurately, are referred to as
machine learning potentials (MLPs). 
MLMD simulation enables accurate and large-scale MD simulations with near DFT accuracy.
Currently, MLMD simulations are extensively utilized for simulating various systems \cite{kobayashi2023machine, nagai2024high,urata2024applications}, including nuclear fuels \cite{kobayashi2022machine, dubois2024atomistic, stippell2024building}.
In this study, we employ MLMD simulations to investigate the high-temperature properties of nuclear materials with 
 (anti-)fluorite structures. 

\begin{figure}[ht]
\begin{center}
\includegraphics[width=1\linewidth]{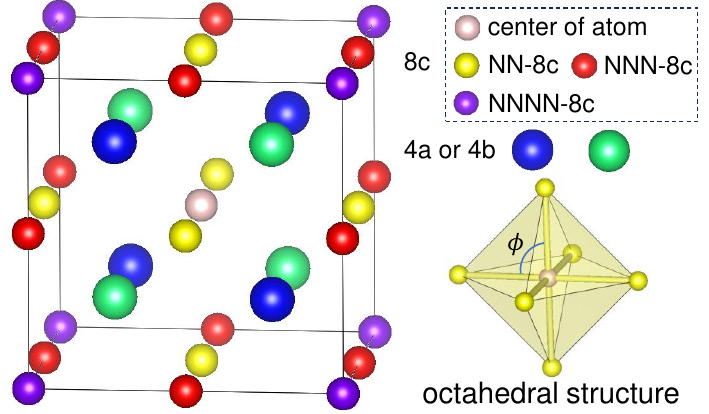}
\end{center}
\caption{
Wyckoff positions of the (anti-)fluorite structure with space group Fm$\bar{3}$m. Thorium and oxygen atoms occupy the 4a (green sphere) and 8c positions (white, yellow, red, and purple spheres) in ThO$_2$, while lithium and oxygen atoms occupy the 8c and 4a positions in Li$_2$O, respectively. The 4b position is an interstitial site. NN-8c, NNN-8c, and NNNN-8c mean the nearest neighboring 8c, next-nearest neighboring 8c, and next-next-nearest neighboring 8c positions from the center of an 8c position. The octahedral structure represented by a yellow polygon denotes the minimal local order for oxygen in ThO$_2$ and lithium in Li$_2$O.
}
\label{Fig1}
\end{figure}

One of the significant phenomena of (anti-)fluorite materials at high temperature is specific heat anomaly, 
known as the $\lambda$-transition, Bredig transition, or pre-melting phase transition \cite{Bredig_transition}.  
In a fluorite structure, the 4a and 8c positions are occupied by cations and anions, respectively (see Fig.\ref{Fig1}). 
Conversely, in an anti-fluorite structure, the 4a and 8c positions are occupied by anions and cations, respectively.
The specific heat anomaly of (anti-)fluorite materials is considered to be attributed to the disorder of 8c position atoms, 
the formation of Frenkel pair like defects \cite{Hutchings_1984, PhysRevLett.52.1238, hull2004superionics} 
at high temperature.
However, the diffusion of mobile atoms is highly cooperative and complex.
The dynamics of the mobile atoms in fluorite materials exhibit string-like atomic diffusion \cite{string1,string2}, which is observed in glass-forming system \cite{donati1998stringlike}. 
The ionic conductivity of fluorite materials exhibits a rapid increase with rising temperature, leveling off at a value indistinguishable from that of the liquid state above the $\lambda$-transition temperature \cite{derrington1973anion, carr1978electrical}. 
The highly cooperative and liquid-like fast atomic diffusion near the $\lambda$-transition temperature complicates the characterization of defect structures in (anti-)fluorite materials within the framework of lattice defects.

In this paper, we investigate the high-temperature properties of ThO$_2$ (fluorite structure) and Li$_2$O (anti-fluorite structure) using MLMD simulations, with a focus on characterizing their heat anomalies from the perspective of local symmetry breaking.
First, we construct MLPs for ThO$_2$ and Li$_2$O. 
Using a farthest-point sampling method \cite{CUR, cersonsky2021improving} and evaluating the uncertainty in the DFT reference data, we generate 1,000 DFT reference data and create MLPs for both ThO$_2$ and Li$_2$O.
The ThO$_2$-MLP created from these 1,000 DFT data achieves almost the same accuracy as the previously developed model \cite{kobayashi2022machine} and well reproduces experimental data.
Similarly, the Li$_2$O-MLP accurately reproduces reported experimental properties, such as lattice constants, thermal expansion coefficients, Frenkel pair formation energy, and melting temperature.
We also discuss the $\lambda$-peak of specific heat capacity and its corresponding temperature for Li$_2$O, which have not been experimentally measured.
Next, using the MLMD trajectories for ThO$_2$ and Li$_2$O, we examine the defective structures contributing to the specific heat anomalies.
We focus on the similarities between the specific heat anomalies of (anti-) fluorite structures and a liquid-liquid phase transition in network-forming liquids such as supercooled water and liquid silica \cite{tanaka2002simple, angell2000water, saika2001fragile}.
These systems also exhibit a $\lambda$-peak of specific heat capacity due to their local symmetry breaking of a local tetrahedral structure.
For ThO$_2$ and Li$_2$O, we formulate the heat anomalies from the perspective of local symmetry breaking within octahedral structures, which represent the minimal locally ordered arrangements of mobile atoms.
The defect structures in ThO$_2$ and Li$_2$O below and above the $\lambda$-transition temperature are characterized by changes in the distribution of the octahedral local order parameter.

\section{Method}
To construct a machine learning potential (MLP), it is necessary to prepare a diverse set of DFT reference data. 
We employed classical MD simulations to generate various structural configurations of ThO$_2$ and Li$_2$O, 
In this paper, all MD simulations are carried out by LAMMPS \cite{Lammps}.
For ThO$_2$, the Cooper, Rushton, and Grimes (CRG) potential was utilized \cite{CRG}, while for Li$_2$O, the Asahi potential was used \cite{Asahi_2014}.
{\it NPT} simulations were performed for ThO$_2$ over a temperature range of 100 K to 5000 K with a 200 K temperature step, and for Li$_2$O from 100 K to 2500 K with a 100 K temperature step. 
The $2\times2\times2$ and $3\times3\times3$ supercell structures (96 and 324 atoms, respectively) were used for the {\it NPT} simulations.
At each temperature step, an {\it NPT} simulation with 100,000 time steps was executed, collecting structural data at intervals of 100 steps. 
Consequently, the structures of ThO$_2$ and Li$_2$O generated through classical MD were each 50,000 structures. 
From these configurations, important structures were selected using a farthest-point sampling (FPS) methodology \cite{CUR,cersonsky2021improving}. 
Initial structures for FPS were selected from perfect crystals at low temperatures.
Subsequent configurations were selected according to the following criteria:
\be
k =
\underset{k \in K} {\operatorname{argmax}} 
(\underset{j \in J} {\operatorname{min}}  |\bo{\Phi}_k - \bo{\Phi}_j |)\,,
\ee
where 
$\bo{\Phi}$ represents structural descriptor, 
$|\cdot|$ is the Euclidean distance, 
$J$ denotes the set of already selected structures and 
$K$ represents the set of structures yet to be selected.
We constructed the structural descriptor of ThO$_2$ and Li$_2$O as 
$
\bo{\Phi}_{m} = 
\frac{1}{N_m}
\bigoplus_{\alpha}
\left(\sum_{i_\alpha }\bo{G}_{i_\alpha }^{(m)}\right)
$, 
where $N_m$ is a number of atoms for $m$-th structure, 
$\alpha$ denotes the chemical species ($\alpha$=O, Th or Li), 
and $\bo{G}_{i_\alpha}$ represents $i_\alpha$-th local atomic environmental descriptor.
The following symmetry functions \cite{BPNN1,BPNN2} were adopted as the local atomic environmental descriptors: 
\be
&& \bo{G}_{i} = \bo{G}_{{\rm R}, i}\oplus \bo{G}_{{\rm A},i}\,, \label{SF} \\
&& G_{{\rm R},i} = \sum_{j}
e^{-\eta_{\rm R}(R_{ij}-R_{\rm s})^2}f_{\rm c}(R_{ij}) \,, \\
&&G_{{\rm A},i} = 2^{1-\xi}\sum_{j\neq i}\sum_{k\neq i,j}
\left(1+\lambda\cos\theta_{ijk}\right)^{\xi} 
e^{-\eta_{\rm A}(R_{ij}^{2}+R_{ik}^{2}+R_{jk}^{2})}\nn\\
&&\qquad\qquad \times f_{\rm c}(R_{ij})f_{\rm c}(R_{ik})f_{\rm c}(R_{jk})\,,
\ee
with the cutoff function
\be
f_{\rm c}(R)=
  \begin{cases}
    0.5\cos\left(\frac{\pi R}{R_{\rm c}}+1\right) \quad {\rm for} \quad R\le R_{\rm c}\\
    0 \hspace{2.6cm} {\rm for} \quad  R_{\rm c} < R
  \end{cases}\,,
\ee
where $R_{ij}$ is the distance between the $i$-th and $j$-th atoms, $\theta_{ijk}$ is the angle formed by line segments between the $i$-$j$-th and the $i$-$k$-th atom bonds.
We utilized the previous parameterization of symmetry functions 
$(R_{\rm c}, \eta_{\rm R}, \eta_{\rm A}, R_{\rm s}, \lambda,$ and $\zeta )$ for ThO$_2$\cite{kobayashi2022machine}\,.
For Li$_2$O, the cutoff radius $R_{\rm c}$ for $G_{{\rm R},i}$ and $G_{{\rm A},i}$ were taken as 6.5 $\text{\AA}$.
The other parameters of symmetry functions for Li$_2$O were selected by CUR decomposition \cite{CUR}.
Using FPS with the structural descriptor $\bo{\Phi}_{m}$, 
we selected 100 configurations from 50,000 structures for both ThO$_2$ and Li$_2$O using FPS.

We conducted DFT calculations for the 100 configurations selected by FPS. 
The Vienna Ab-initio Simulation Package (VASP) \cite{VASP1, VASP2} was utilized for these calculations. 
In all calculations, the projector-augmented wave (PAW) method \cite{PAW} was employed, with an energy cutoff of 500 eV. 
The strongly constrained and appropriately normed (SCAN) meta-GGA exchange-correlation functional \cite{SCAN} was used, since MLP based on the SCAN functional is suitable to reproduce high-temperature thermal properties of ThO$_2$ \cite{kobayashi2022machine}.
We trained Behler-Parrinello neural networks type MLPs on 100 DFT reference data sets for ThO$_2$ and Li$_2$O using the n2p2 code \cite{n2p2}.
The symmetry function defined in Eq.(\ref{SF}) was used as input of the neural network.
The neural network architecture consisted of two hidden layers, each with 30 nodes, and hyperbolic tangent activation functions. 

By identifying structures with high uncertainty in the MLP outputs for various structures and incorporating these structures into the DFT reference data, the overall quality of the reference data can be improved \cite{Botu2015, PhysRevLett.114.096405, Gastegger2017, Jacobsen2018, Li2019, tan2023single}.
Therefore, using the MLPs created from the 100 DFT reference data sets, 
we again conducted NPT simulations over a temperature range of 300 to 4500 K (300 to 2500 K) for ThO$_2$ (Li$_2$O), to generate various structures.  
The uncertainty of the MLP outputs was assessed by evaluating the variance of the outputs from multiple MLPs.
We divided a set of the 100 DFT reference data, $S$, into five non-overlapping datasets, $S_{l=1,\cdots,5}$, 
and created five MLPs trained by the 80 DFT reference data from $S \setminus S_{l}$.
Using ensemble of MLPs outputs, 
we preferentially selected structures that exhibited a large maximum force prediction variance. 
The maximum force prediction variance was defined as 
$
{\rm Var} (\bo{F}^{(m)}) \equiv 
\max_{i_m} 
\langle |\bo{F}_{i_m}^{(m)} - \langle \bo{F}_{i_m}^{(m) }\rangle|^2 \rangle
$, 
where $\bo{F}_{i_m}^{(m)}$ denotes the force acting on the $i_m$-th atom in the $m$-th structure, and the brackets $\langle \cdot \rangle$ represent the ensemble average for outputs of five MLPs.
We also selected the structures that had the symmetry functions in the extrapolation region to ensure that the predictions of the MLP remain as interpolative as possible during MLMD simulations.
The number of extrapolated symmetry functions was calculated as 
$
N_{\rm E}^{(m)} = \sum_{i_m}^{N_m} \sum_{d}^{D}\left[ 1-H_{S}(G_{i,d}^{(m)}) \right] \,,
$
where $N_m$ is a number of atoms for $m$-th structure, 
$D$ denotes the dimension of the symmetry function, 
$H_{S}(G_{i,d})$ is a step function indicating whether $G_{i,d}$ is within the symmetry function range of the DFT reference data, $S$.
We selected 900 structures with a large maximum force prediction variance and a large number of extrapolations, incorporating these structures into the DFT reference data.
The total numbers of the reference data for ThO$_2$ and Li$_2$O used to create MLPs were 1000 structures.

We also created validation data to evaluate the root mean square error (RMSE) of the MLPs.
NPT simulations with the MLPs were conducted at temperatures of 300, 1300, 2300, 3300, and 4300 K for ThO$_2$, and 300, 800, 1300, 1800, and 2300 K for Li$_2$O.
From the trajectory at each temperature, we randomly selected 20 structures and calculated the DFT energy and forces for these structures as validation data.
The resulting RMSEs for the ThO$_2$- and Li$_2$O-MLPs are summarized in Table \ref{tab:table1}.
For the solid phase, the RMSEs of the ThO$_2$- and Li$_2$O-MLPs are below 1 meV/atom for energies and 0.12 eV/$\text{\AA}$ for forces. 
While the RMSEs for the liquid phase are higher than those for the solid phase, the values are comparable to typical RMSEs reported in previous studies \cite{BPNN1, Khaliullin2011, Morawietz2016, aenet, Li2019}.

\begin{table}[ht]
\small
    \caption{RMSE of ThO$_{2}$- and Li$_2$O-MLP}
\label{tab:table1}
\begin{tabular*}{0.5\textwidth}{@{\extracolsep{\fill}}lccccc}
\hline
\hline
 ThO$_2$-MLP           &    &  &  &  & \\
\hline
number of data  & 20   & 20 & 20  & 20  & 20 \\
phase  & solid   & solid & solid & solid & liquid\\
 T [K]           & 300   & 1300  & 2300  & 3300  & 4300  \\
RMSE for energy        & 0.109   & 0.367 & 0.997 & 0.851 & 3.514 \\
 {[}meV/atom{]}           &   &  &  &  &  \\
RMSE for force          & 0.015   & 0.035 & 0.063 & 0.119 & 0.190 \\
 {[}eV/\AA{]}           &   &  &  &  &  \\
\hline
\hline
Li$_2$O-MLP           &    &  &  &  & \\
\hline
number of data  & 20   & 20 & 20  & 20  & 20 \\
phase  & solid   & solid & solid & solid & liquid\\
 T [K]           & 300  & 800 & 1300 & 1800 & 2300 \\
RMSE for energy          & 0.092   & 0.222 & 0.287 & 0.457 & 4.879 \\
 {[}meV/atom{]}           &   &  &  &  &  \\
RMSE for force          & 0.028   & 0.027 & 0.031 & 0.041 & 0.245 \\
 {[}eV/\AA{]}           &   &  &  &  &  \\
\hline
\end{tabular*}
\end{table}

\section{Results and Discussion}

\subsection{Thermal properties of ThO$_2$ and Li$_2$O computed by MLMD}

\begin{table}[ht]
\small
    \caption{
Comparison of the physical properties of ThO$_2$ computed in this study with previous MLMD study and experiments.
The lattice constant $L(T)$ at 300 K,
the averaged coefficient of linear thermal expansion (ACLTE) in the range from 300 to 1600K,
the onset temperature of heat capacity anomaly $T_{\rm a}$,
$\lambda$-transition tmperature $T_{\rm c}$, and
melting point
are summarized.}
\label{tab:table2}
\begin{tabular*}{0.5\textwidth}{@{\extracolsep{\fill}}crrr}
\hline
       & \multicolumn{2}{c}{MLMD}  & Exp.  \\
       & this study  &previous study \cite{kobayashi2022machine}&   \\
       & (1,000 data)  &(7,007 data)&   \\
\hline
 $L(300{\rm K})$ [\AA]  &5.617  &5.624      & 5.592 \cite{lattice_constant}  \\
ACLTE {[}$10^{-6}$K${}^{-1}${]}&10.02  &9.71   & $9.5$ \cite{CLTE1} \\ 
 &  &   & $9.67$ \cite{CLTE2}\\ 
 &  &   & $11.07$ \cite{IAEA}\\ 
 $T_{\rm a}$ [K]   &2460  &2460 & \\
 $T_{\rm c}$ [K]   &3200  &3200 & $2950$ \cite{transition_temp1}\\
  &  & & $3090$ \cite{transition_temp2}\\
 $T_{\rm m}$ [K]  &3610  &  3610-3620 & $3651$ \cite{transition_temp1}\\
\hline
\end{tabular*}
\end{table}

\begin{figure}[ht]
\begin{center}
\includegraphics[width=1\linewidth]{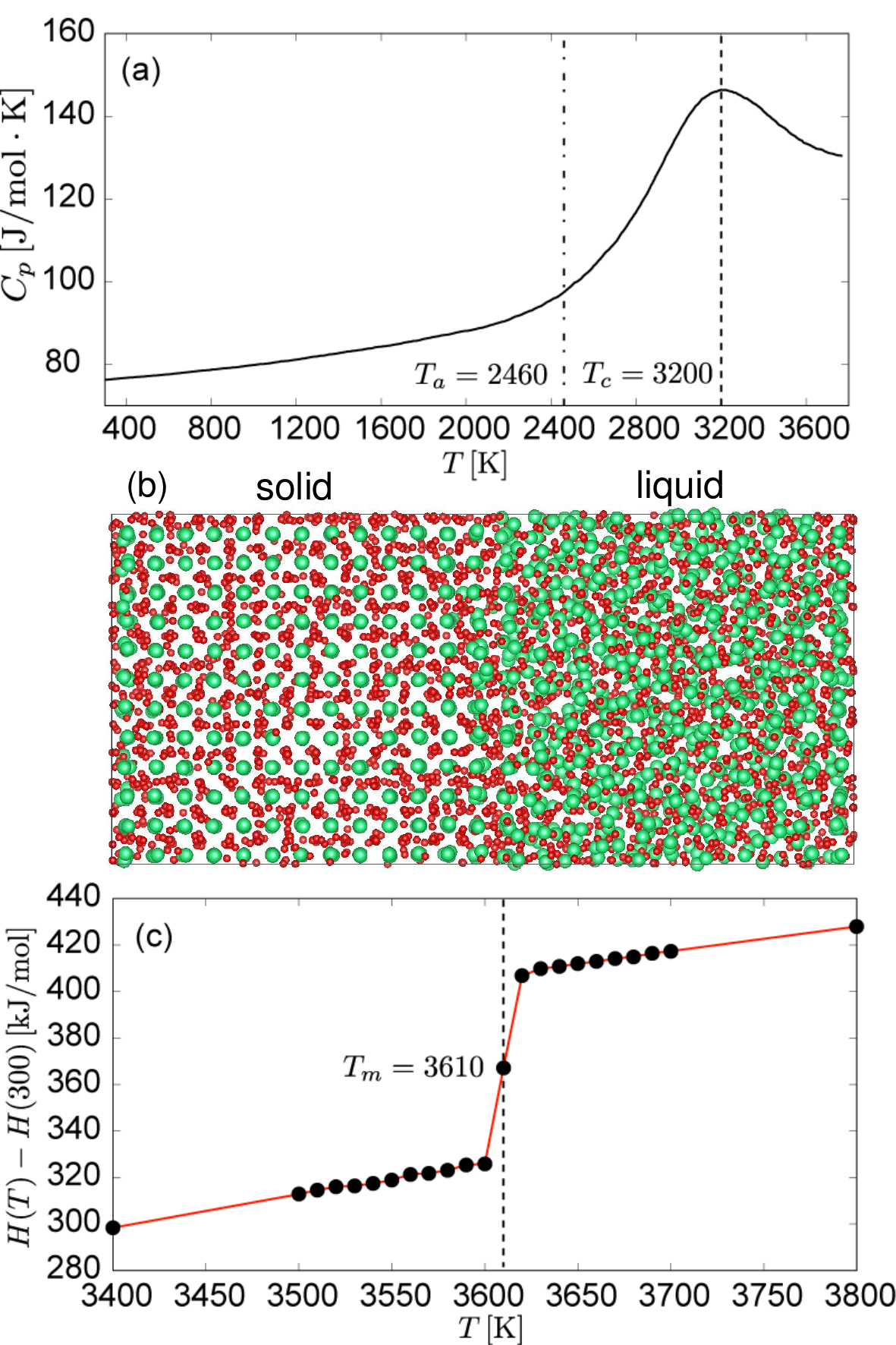}
\end{center}
\caption{
The thermal properties of ThO$_2$ obtained by MLMD.
(a) Molar specific heat capacity of ThO$_2$ computed by MLMD.
The dashed line denotes the peak position of $C_p$ ($\lambda$-peak temperature $T_c$).
The dashed-dot line represents the onset temperature of heat capacity anomaly $T_{\rm a}$,
(b) $6\times6\times12$ supercell including solid (left) and liquid phase (right) for two-phase simulation.
Green and red balls represent thorium and oxygen atoms, respectively
(c) Temperature dependence of the enthalpy of ThO$_2$ obtained by the two-phase simulation approach using MLMD.
}
\label{Fig2}
\end{figure}

First, we demonstrate that the present ThO$_2$-MLP created with 1,000 data has sufficient performance comparable to the MLP previously developed with 7,007 data set \cite{kobayashi2022machine}.
We conducted MLMD-{\it NPT} simulations for ThO$_2$ using a 6$\times$6$\times$6 supercell (2592 atoms), with a time step of 2 fs and a total simulation time of 200 ps.
Table \ref{tab:table2} summarizes the thermal properties of ThO$_2$ computed by the present ThO$_2$-MLP.
The lattice constant $L(T)$ at $T=300$ K obtained by the present MLMD agrees well with the previous MLMD calculation and experimental data \cite{lattice_constant}.
The averaged coefficient of linear thermal expansion (ACLTE), defined by
$
\text{ACLTE} = \frac{1}{T_2-T_1}\int_{T_1}^{T_2} \frac{1}{L(T)}\frac{d L(T)}{d T} dT
$,
in the range from $T_1=300$ K to $T_2 = 1600$ K, is also comparable with values reported in the previous study and experimental data \cite{CLTE1, CLTE2, IAEA}.
Figure \ref{Fig2}(a) shows the molar specific heat capacity, $C_{\rm p}$, calculated by MLMD.
The specific heat capacity exhibits an anomaly at high temperatures, attributed to the disorder of oxygen atoms.
The onset temperature of the specific heat anomaly, $T_{\alpha}$, is characterized as the temperature at which the change in entropy, $dS/dT = C_{\rm p}/T$, reaches a minimum value \cite{eapen2017entropic}.
The peak position of $C_{\rm p}$ is defined as the $\lambda$-transition temperature.
Both $T_\alpha$ and $T_{\rm c}$ are consistent with previous MLMD results, and $T_{\rm c}$ is comparable to experimental observations.
Finally, the melting point was evaluated using a two-phase simulation approach.
A 6$\times$6$\times$12 supercell (5184 atoms), containing both solid and liquid phases, was used as the initial configuration.
MLMD-{\it NPT} simulations were conducted at temperatures ranging from 3400 K to 3800 K, with a total simulation time of 500 ps.
The melting point, determined from the temperature corresponding to the enthalpy jump, is in excellent agreement with prior results and experimental data (see Fig.\ref{Fig2}(b) and Table \ref{tab:table2}).
In conclusion, the present ThO$_2$-MLP, developed with a dataset of 1,000 configurations, is shown to be sufficiently reliable when compared to previous MLMD results and experimental data.
Similarly, the Li$_2$O-MLP, constructed with the same dataset size, is expected to yield reliable predictions for the thermal properties of Li$_2$O.

\begin{figure}[ht]
\begin{center}
\includegraphics[width=1\linewidth]{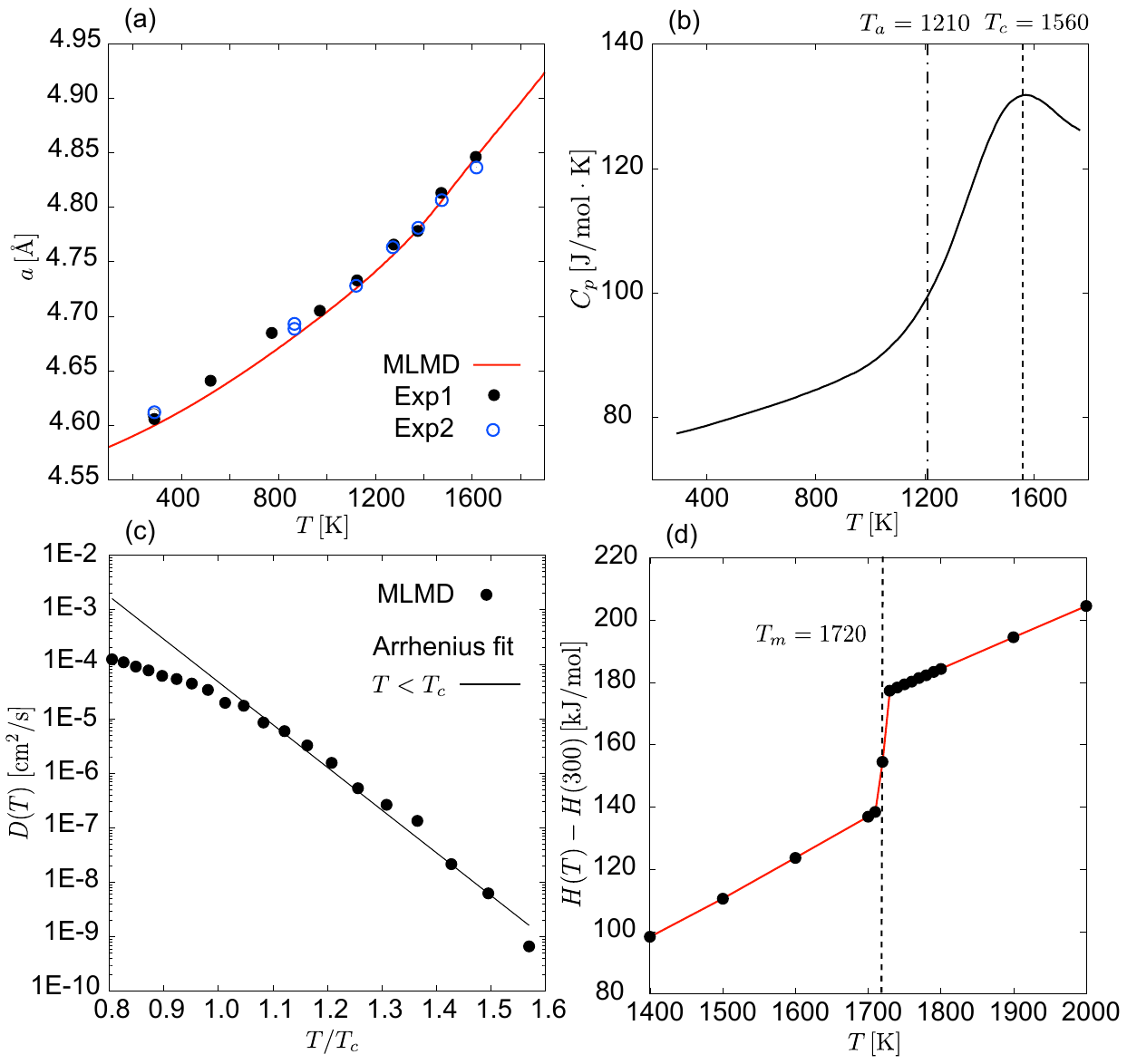}
\end{center}
\caption{
The thermal properties of Li$_2$O obtained by MLMD.
(a) Temperature dependence of the lattice constant. The solid line represents the results obtained by MLMD, while the filled and open circles represent experimental results (Exp1 \cite{HULL_1988}, Exp2 \cite{TWD_Farley_1991}).
(b) Molar specific heat capacity ($C_p$) computed by MLMD.
The dashed line denotes the peak position of $C_p$ ($\lambda$-peak temperature $T_c$).
The dashed-dot line represents the onset temperature of heat capacity anomaly $T_{\rm a}$,
(c) Arrhenius plot of the self-diffusion constant for lithium atom (filled circles).
The solid and dashed lines are the results of Arrhenius fitting below and above the transition temperature $T_c$.
(d) Temperature dependence of the enthalpy obtained by two-phase simulation approach using MLMD.
}
\label{Fig3}
\end{figure}

Next, we investigate the thermal properties of Li$_2$O using MLMD simulations.
MLMD-{\it NPT} simulations were conducted for Li$_2$O with a time step of 1 fs.
The system size and total simulation time were the same as those used in the MLMD calculations for ThO$_2$.
The lattice constant obtained from the MLMD simulations shows good agreement with the reported experimental data over a wide temperature range \cite{TWD_Farley_1991, Juza_1957, HULL_1988} (see table \ref{tab:table3} and Fig.\ref{Fig3}(a)).
Additionally, the ACLTE computed in the range from 300 to 1280 K also agrees with experimental value \cite{KURASAWA1982334}.
The onset temperature of heat capacity anomaly, $T_{\rm a}$, and $\lambda$-peak temperature, $T_c$, obtained from specific heat capacity, $C_p$, in Fig.\ref{Fig3}(b) are 1210 and 1560 K, respectively.
However, available experimental data on specific heat capacity are restricted to temperatures below 1125 K \cite{shomate1955high, barin2013thermochemical, tanifuji1979heat}, and a $\lambda$-peak in specific heat capacity was not observed experimentally.
On the other hand, neutron diffraction experiments \cite{TWD_Farley_1991} indicate that the onset temperature of lithium disorder, characterized by a rapid increase in the fraction of lithium atoms leaving their regular sites, occurs at approximately 1200 K.
This onset temperature of lithium disorder reported by neutron diffraction experiments corresponds to the onset temperature of the heat capacity anomaly, $T_{\rm a} = 1210$ K, computed in the present MLMD calculation.
To calculate the self-diffusion constant, we conducted NVE simulations with a 0.5 fs time step and a total simulation time of 200 ps at various temperatures using the lattice constants previously obtained in the {\it NPT} ensemble.
Figure \ref{Fig3}(c) shows the Arrhenius plot of the self-diffsuion constant, $D(T)$, of lithium atom obtained from the slope of the mean square displacement.
The Arrhenius plot of $D(T)$ exhibits the downward bending above the $T_{\rm c}$.
The activation energy of lithium diffusion, $E_{{\rm AE}}$, in the range 1000 K $\le  T < T_{\rm c}$ using the Arrhenius fitting, $D(T) = D_{0} \exp(-E_{\rm AE}/k_{\rm B}T)$, become 2.44 eV, which well agree with experimental value, $2.52$ eV \cite{oishi1979self}.
We also evaluated the formation energy of a Frenkel pair defect, consisting of a lithium interstitial at a 4b position and a vacancy at an 8c position, using structural optimization.
The Frenkel pair defect formation energy, $E_{\rm FP}$, closely matches experimental data, as shown in Table \ref{tab:table3}.
The melting point obtained from the present MLMD simulations is in excellent agreement with the experimental melting point (1711 K), as shown in Fig.\ref{Fig3} and Table \ref{tab:table3}.

The results demonstrate that the MLMD simulations using the Li$_2$O-MLP accurately reproduce the thermal properties of Li$_2$O.
Although the $\lambda$-peak in specific heat capacity of Li$_2$O is not reported experimentally, the $\lambda$-peak temperature, $T_c$, obtained by MLMD is likely reasonable, as the MLMD results agree well with other experimental data, including the melting point.
However, it may be difficult to observe the $\lambda$-peak in the specific heat capacity of Li$_2$O clearly in experiments.
The melting point and $\lambda$-peak temperature predicted by the MLMD simulation are very close, and the $\lambda$-peak may be hidden by the jump of specific heat capacity due to the melting of Li$_2$O.

\begin{table}[ht]
\centering
    \caption{Computed and experimental properties of Li$_2$O. 
The lattice constant $L(T)$ at 300 K,
the ACLTE in the range from 300 to 1280K,
the onset temperature of heat capacity anomaly $T_{\rm a}$,
$\lambda$-peak tmperature $T_{\rm c}$,
the activation energy of lithium diffusion $E_{{\rm AE}}$ in the range 1000 K $\le T < T_{\rm c}$,
the Frenkel pair formation energy $E_{{\rm FP}}$,
and
the melting point $T_{\rm m}$
are summarized.
}
\label{tab:table3}
\begin{tabular*}{0.5\textwidth}{@{\extracolsep{\fill}}lrr}
\hline
            & MLMD  & Exp.  \\
\hline
 $L(300{\rm K})$ [\AA]  &4.601         & 4.610\cite{TWD_Farley_1991}, 4.628\cite{Juza_1957}\\
ACLTE [$10^{-6}$K${}^{-1}$]  &34.01    & 33.6\cite{KURASAWA1982334}\\ 
 $T_{\rm a}$ [K]   &1210  &   1200 \cite{TWD_Farley_1991}\\
 $T_{\rm c}$ [K]   &1560  &          \\
 $E_{{\rm AE} }$ [eV]     &2.44  &    2.52 \cite{oishi1979self}\\
$E_{{\rm FP}}$ [eV]     &2.29  &2.10\cite{TWD_Farley_1991}, 2.6\cite{Chadwick1991}  \\
$T_{\rm m}$ [K]  &1720  &  1711 \cite{ORTMAN1983}\\
\hline
\end{tabular*}
\end{table}

\subsection{Characterization of the specific heat anomalies of ThO$_2$ and Li$_2$O by local order parameter}

We have demonstrated that MLMD simulations are reliable for discussing the high-temperature properties of ThO$_2$ and Li$_2$O.
Using the trajectories obtained from MLMD, we analyze the specific heat anomalies of ThO$_2$ and Li$_2$O.
We focus on the similarities between the specific heat anomalies of (anti-)fluorite structures and 
the liquid-liquid phase transitions in network-forming liquids, such as supercooled water (H$_2$O) 
and liquid silica (SiO$_2$) \cite{tanaka2002simple, angell2000water, saika2001fragile}.
These network-forming liquids also exhibit a $\lambda$-peak in specific heat capacity and bending in the Arrhenius plot of the self-diffusion constant at the transition temperature. 
The liquid-liquid transitions in these network-forming liquids are characterized by local symmetry breaking in their atomic arrangements.
The local tetrahedral structures formed by oxygen atoms in supercooled water and silicon atoms in liquid silica play a key role in defining their liquid-liquid transitions \cite{errington2001relationship, kumar2009tetrahedral, geske2016fragile, shi2018impact}. 
Drawing an analogy between liquid-liquid phase transitions and the specific heat anomalies of (anti-)fluorite structures, we focus on the local atomic configurations of oxygen atoms in ThO$_2$ and lithium atoms in Li$_2$O.
The octahedral structure, consisting of six nearest-neighboring atoms surrounding a central atom, can be considered the minimal local ordered structure of the sublattice in (anti-)fluorite structures (see Fig.\ref{Fig1}). 
To quantify this local order, we define the local octahedral order parameter as 
\be
Q_i = 1 - \frac{1}{8} \sum_{j\in \text{6NN$_i$}}\sum_{k\in \text{4NN}_j} \cos^2 \phi_{ijk}\,, \lb{eq:ORDER}
\ee
where $\phi_{ijk}$ denotes the internal angle of the octahedron. 
The summation over $j \in \text{6NN}_i$ accounts for the six nearest neighboring atoms relative to the $i$-th atom, while the $k \in \text{4NN}_j$ represents the four nearest neighboring atoms within the 6NN$_i$ group from the $j$-th atom. 
The order parameter $Q_i$ equals one when the six nearest neighboring atoms of the $i$-th atom form an ideal octahedral structure, characterized by internal angles of 90 degrees. 
The expectation value of the order parameter becomes zero,  $\langle Q_i \rangle_{\rm rand}=0$, if the internal angles, $\phi_{ijk}$, are completely random.

\begin{figure*}[ht]
\begin{center}
\includegraphics[width=1\linewidth]{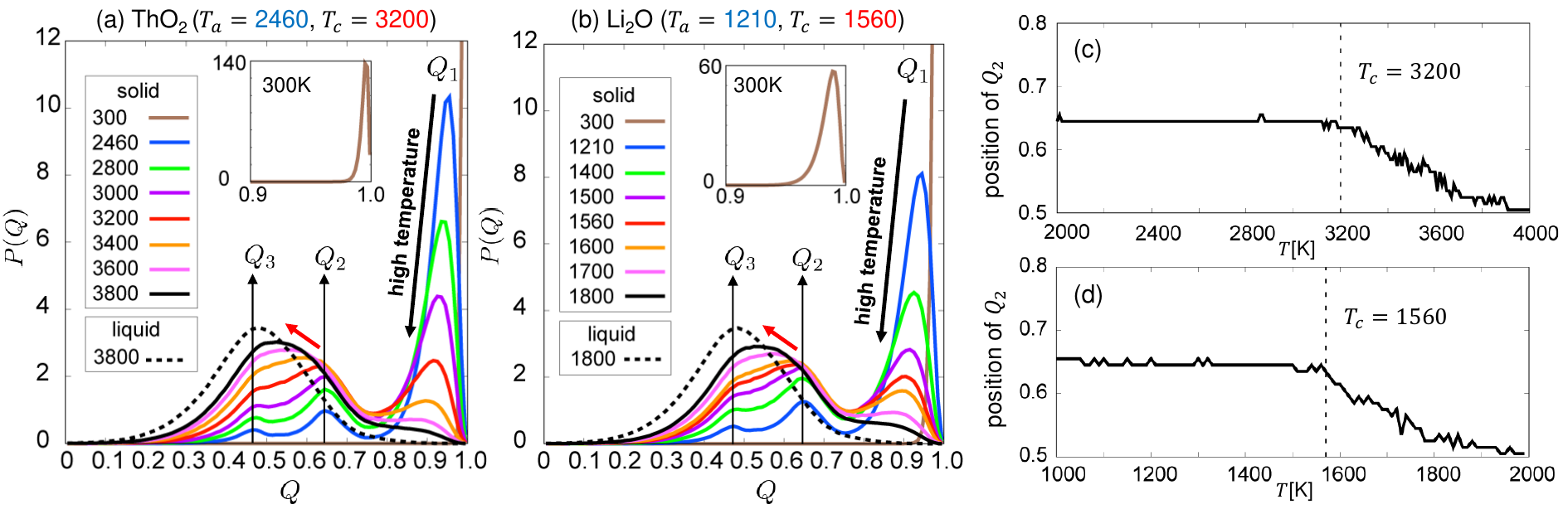}
\end{center}
\caption{
(a) and (b): The distribution of the local octahedral order parameter for the oxygen atoms in ThO$_2$ and lithium atoms in Li$_2$O, 
$P(Q)$, for ThO$_2$ 
and Li$_2$O, respectively. 
The solid lines represent $P(Q)$ for the solid phase, while the dashed lines denote the liquid phase. 
The insets show the results at 300 K for clarity. 
(c) and (d) show the change of the position of $Q_2$-peak in $P(Q)$ for ThO$_2$ and Li$_2$O, respectively. 
}
\label{Fig4}
\end{figure*}

\begin{figure*}[ht]
\begin{center}
\includegraphics[width=1\linewidth]{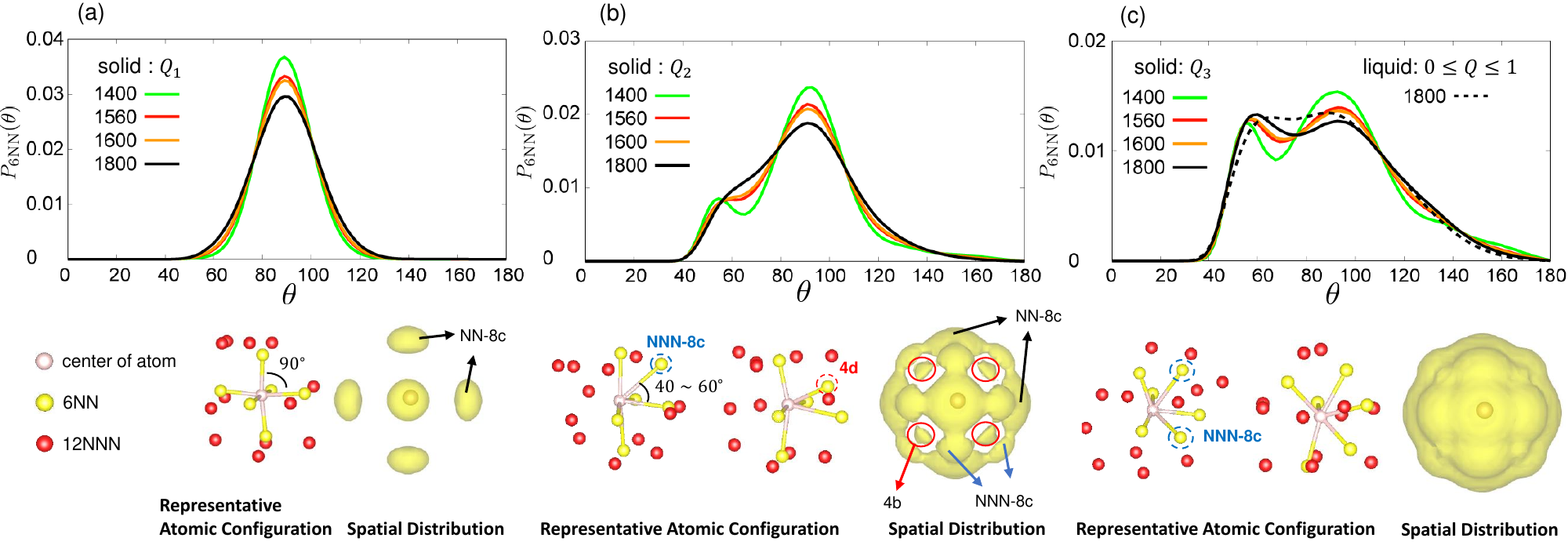}
\end{center}
\caption{
The angle distribution of the six nearest neighboring (6NN) lithium atoms, $P_{\rm 6NN}(\theta)$, relative to a central lithium atom in Li$_2$O.
(a)-(c) show $P_{\rm 6NN}(\theta)$ corresponding to the local atomic arrangements with octahedral order parameter values at the following ranges:
$Q_1 (0.755 < Q \le 1) $,
$Q_2 (0.535 < Q \le 0.755)$,
$Q_3 (Q\le 0.535)$, respectively. 
For comparison, the angle distribution for the liquid phase is also shown in panel (c) (dashed line). Here, the angle distribution for the liquid phase is plotted across the range, $0\le Q \le 1$. 
The lower panels illustrate representative atomic configurations and spatial distributions for each $Q$-region.
The white, yellow, and red balls represent the center of the lithium atom, the six nearest neighboring (6NN) lithium atoms, and
the twelve next-nearest neighboring (12NNN) lithium atoms, respectively.
The spatial distributions denote the distributions of 6NN lithium atoms relative to the center of the lithium atom.
The configurations and spatial distributions are obtained from the MLMD trajectory of lithium for Li$_2$O at 1600 K.
}
\label{Fig5}
\end{figure*}

We calculated the distribution of the local octahedral order parameter, $P(Q)$, from the trajectories of the MLMD,  as shown in Fig.\ref{Fig4}(a) and \ref{Fig4}(b). 
At 300 K, $P(Q)$ exhibits a sharp peak, $Q_1\simeq 1$, indicative of the perfect local octahedral order of oxygen (lithium) atoms in ThO$_2$ (Li$_2$O).
As the temperature increases, the height of the $Q_1$-peak decreases due to the disruption of the octahedral atomic arrangement caused by large thermal vibrations.
Additionally, new peaks representing defective local structures emerge in the low-$Q$ region at higher temperatures. 
The low $Q$-peaks becomes predominant near the $\lambda$-transition temperature, $T_c$. 
This behavior closely resembles the change in the local tetrahedral order parameter for the liquid-liquid transition of super-cooled water and liquid silica \cite{kumar2009tetrahedral,geske2016fragile,shi2018impact,shi2019distinct}. 
Comparing the temperature change of the local tetrahedral order parameter in these network-forming liquids, 
the distributions of the local octahedral order parameter in the low $Q$ region show two peaks, $Q_2$ and $Q_3$, in the present (anti-)fluorite systems.  
The position of the $Q_3$-peak is nearly identical to the peak position of the liquid state (see dashed lines in Fig.\ref{Fig4}(a) and Fig.\ref{Fig4}(b)). 
Interestingly, two peaks, $Q_2$ and $Q_3$,  merge into a single single low $Q$-peak at $T_{\rm c}$, 
indicating the qualitative changes of the defect structures in ThO$_2$ and Li$_2$O associated with the $\lambda$-transition. 

To elucidate the significance of each peak, 
we computed the angle and atomic spatial distributions of the six nearest neighboring (6NN) atoms around a central atom with $Q$ values belonging to the $Q_1$-, $Q_2$-, and $Q_3$-peak regions.
The $Q_1$-, $Q_2$- and $Q_3$-peak regions were assigned on the basis of the local minimum values of $P(Q)$ at the onset temperature of heat capacity anomaly, $T_{\rm a}$. 
The $Q_1$-peak region is $0.755<Q\le 1$, the $Q_2$ peak region is $0.535<Q\le 0.755$, and the $Q_3$-peak region is $Q \le 0.535$, which are common for both ThO$_2$ and Li$_2$O.
Figure \ref{Fig5}(a)-(c) show the angle distributions belonging to the $Q_1$-, $Q_2$- and $Q_3$-peak regions for Li$_2$O.
A single peak is observed at 90 degrees in the angle distribution for the $Q_1$ peak, representing the ideal octahedral structure, as shown in Fig.\ref{Fig5}(a). 
For the angle distribution of the $Q_2$ peak, the distribution more broadens, and a small peak appears between 40 and 60 degrees. 
This small peak is attributed to the atomic arrangement where the 6NN atoms partially occupy either the next-nearest neighboring 8c (NNN-8c) positions or the interstitial 4b positions (see the lower panel of Fig.\ref{Fig5}(b)).
This is because, if a 6NN atom partially occupies a position near NNN-8c or 4b, the angle becomes approximately 40 to 60 degrees. 
Indeed, the occupancy of both the NNN-8c and interstitial 4b positions are confirmed in the atomic spatial distribution in Fig.\ref{Fig5}(b).
Thermal vibrations at high temperatures induce significant anharmonicity, disrupting the octahedral arrangement of the 6NN atoms. 
The presence of 4b occupancies confirms that the $Q_2$-peak reflects a lattice defect-like local structure 
(a Frenkel pair defect-like local structure). 
In the angle distribution for the $Q_3$ peak, the peak between 40 and 60 degrees accounts for a more significant proportion than the distribution of the $Q_2$ peak. 
This increase in the peak corresponds to cases where multiple NNN-8c positions are included among the 6NN atoms (see the lower-left panel of Fig.\ref{Fig5}(b)) or to disordered local structures that are difficult to assign to specific Wyckoff positions (see the lower-middle panel of Fig.\ref{Fig5}(b)). 
The atomic spatial distribution for the $Q_3$-peak also has local maximums at the NN-8c and NNN-8c position, similar to the $Q_2$-peak distribution, but the 4b position does not become a local maximum. 
Therefore, the $Q_3$ peak is considered to represent a local defect structure qualitatively different from the lattice defect-like local structure of the $Q_2$ peak.
The position of the $Q_3$ peak in $P(Q)$ is nearly identical to the peak position of the liquid state  (see dashed lines in Fig.\ref{Fig4}(b)), and its angular distribution, $P_{\rm 6NN}(\theta)$, closely resembles that of the liquid phase (see dashed lines in Fig.\ref{Fig5}(c)).
Thus, the $Q_3$-peak represents the disordered local structure resembling the local structure of the liquid phase.

\begin{figure}[ht]
\begin{center}
\includegraphics[width=1\linewidth]{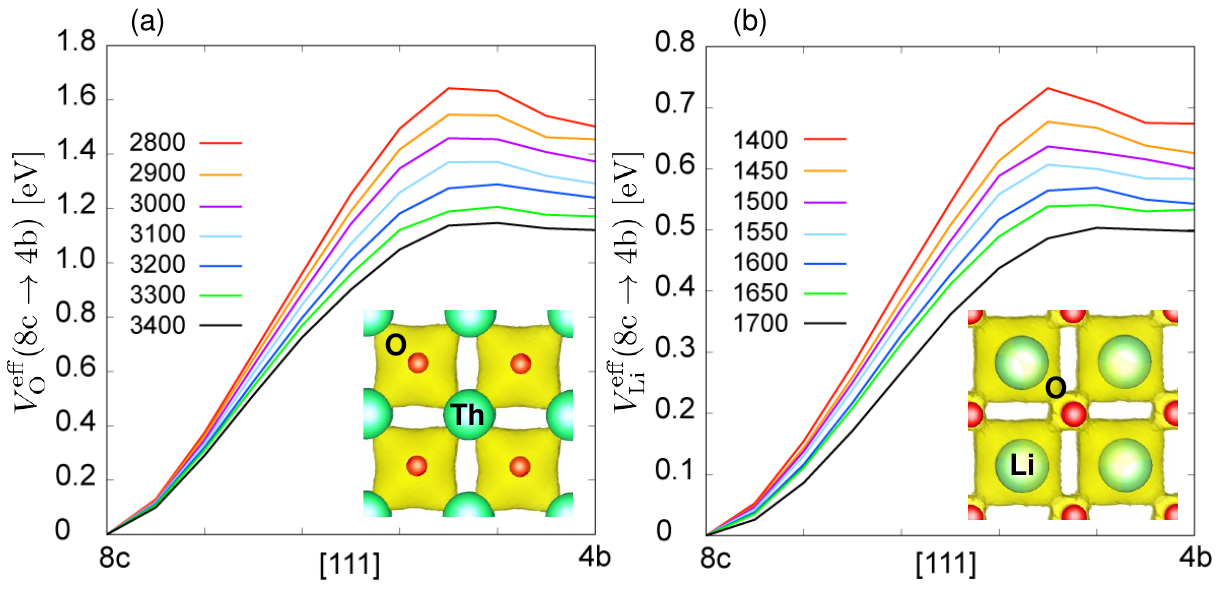}
\end{center}
\caption{
(a) and (b) represent effective one-particle potential energies of oxygen and lithium atoms in ThO$_2$ and Li$_2$O for the $[111]$ direction from 8c to 4b position.
}
\label{Fig6}
\end{figure}

We have shown that the mobile atoms in ThO$_2$ and Li$_2$O have three characteristic local structures at high-temperature: 
the octahedral local structure ($Q_1$), 
the lattice defect-like local structure ($Q_2$), and 
the liquid-like local structure ($Q_3$).
The $Q_2$ and $Q_3$ peaks merge into a single low-$Q$ peak at $T_{\rm c}$, indicating a qualitative change in the mobility of atoms in ThO$_2$ and Li$_2$O. 
We computed the one-particle atomic distribution of mobile atoms, 
\be
p_{\rm A}(\bo{x})=
\frac{1}{N_{\rm A}}
\langle
\sum_{i}^{N_{\rm A}}\delta(x-x_i)
\rangle
\,,
\ee
where 
$\langle\cdot\rangle$ means thermal average, 
and 
A = O, Li. 
From the $p_{\rm A}(\bo{x})$, we calculate the effective one-particle effective potentials for the $[111]$ direction from 8c to 4b position as 
\be
V_{\rm A}^{\rm (eff)}({\rm 8c}\to{\rm 4b}) \equiv -k_{\rm B}T\ln
\left(
\frac{p_{A}(\bo{x}={\rm 8c}\to{\rm 4b})}{p_{A}(\bo{x}={\rm 8c})}
\right)\,,
\ee
where $k_{\rm B}$ is boltzmann constant. 
As shown in Fig.\ref{Fig6}.(a) and (b), 
$
V_{\rm A}^{\rm (eff)}({\rm 8c}\to{\rm 4b})
$ has a local minimum at the 4b position at low temperature, while the local minimum disappears near $T_{\rm c}$. 
The 4b position does not become a metastable state above $T_{\rm c}$, 
indicating that mobile atoms no longer stagnate in the interstitial 4b position.
This result corresponds to the fusion of the $Q_2$-peak (lattice defect-like local structure) and 
the $Q_3$-peak (liquid-like local structure) above $T_{\rm c}$. 
The sub-lattice formed by mobile atoms melts and 
the defective local structure becomes close to the liquid-like structure above $T_{\rm c}$. 
The bending of the self-diffusion constant of mobile atoms at $T_{\rm c}$ is considered due to the transition of the local structure from the lattice defect-like local structure to the liquid-like one above $T_{\rm c}$.

\section{Conclusion}
We have performed MLMD simulations to explore the high-temperature properties of ThO$_2$ and Li$_2$O, which possess fluorite and anti-fluorite structures, respectively.
Our results show that the MLMD simulations accurately reproduce the thermal properties reported in experiments.
The MLMD simulations predicted a $\lambda$-transition temperature of 1560 K for Li$_2$O, a value that has not been experimentally reported.
The computed $\lambda$-transition temperature is considered reasonable since the MLMD results well reproduced the other reported experimental data.
However, the predicted $\lambda$-transition temperature is close to the experimental melting point of 1711 K.
This result implies that observing a distinct $\lambda$-peak in specific heat capacity experimentally may be difficult due to the jump in specific heat capacity associated with the melting of Li$_2$O.

Using the MLMD trajectories, we investigated the heat anomalies of ThO$_2$ and Li$_2$O from the perspective of local symmetry breaking, a concept proposed to explain the liquid-liquid transition in network-forming liquids.
We introduced the local octahedral order parameter and analyzed how its distribution changes with increasing temperature.
The changes in these distributions with rising temperature closely resemble the changes observed in the distribution of the tetrahedral order parameter in supercooled water and liquid silica.
The local octahedral order parameter revealed two distinct defective peaks below the transition temperature.
The first peak corresponds to a lattice defect-like local structure, while the second corresponds to a liquid-like local structure.
These two peaks merge into a single peak at $\lambda$-transition temperature,
and mobile atoms no longer stagnate in interstitial 4b positions.
The sub-lattice character of mobile atoms disappears, and the liquid-like local structure becomes dominant above the transition temperature.

\section*{Acknowledgements}
K.K. was supported by JSPS KAKENHI Grants Number 24K08574.
H.N. was supported by JSPS KAKENHI Grants Number 23K04637.
This calculations were mainly performed on the supercomputing system HPE SGI8600 in the Japan Atomic Energy Agency. We would thank all staff members of CCSE for computational support.
We also partially used the computational resources of Fujitsu PRIMERGY CX400M1/CX2550M5 (Oakbridge-CX), Fujitsu PRIMEHPC FX1000, FUJITSU PRIMERGY GX2570 (Wisteria/BDEC-01), and “mdx: a platform for the data-driven future” supported by “Joint Usage/Research Center for Interdisciplinary Large-scale Information Infrastructures” in Japan (Project ID: jh230069).
The crystal structures were drawn with VESTA\,\cite{VESTA}.

\bibliography{reference}

\end{document}